\documentclass[11pt,fleqn]{article}

\usepackage{amsmath,amssymb,amsthm,enumerate,graphics,epsfig,cite}

\newcommand{\todo}[1][\null]{\ensuremath{\clubsuit}}

\newcounter{tbn}

\newcounter{mcasenum}

\newtheorem{theorem}{Theorem}

\newtheorem*{proposition*}{Proposition}
{\theoremstyle{definition}

\newtheorem{remark}{Remark}
}

\begin{document}
\begin{center}
\par\noindent{\LARGE\bf
 Group Classification of\\  Variable Coefficient $\boldsymbol{K(m,n)}$ Equations
\par}

{\vspace{5mm}\par\noindent\large Kyriakos Charalambous$^{\dag 1}$, Olena Vaneeva$^{\ddag 2}$ and Christodoulos Sophocleous$^{\dag 3}$
\par\vspace{5mm}\par}
\end{center}
\vspace{-5mm}
{\par\noindent\it
${}^\dag$\ Department of Mathematics and Statistics, University of Cyprus, Nicosia CY 1678, Cyprus\\[1ex]
${}^\ddag$\ Institute of Mathematics of NAS of Ukraine, 3 Tereshchenkivska Str., Kyiv-4, 01601 Ukraine
}

\vspace{2mm}
{\par\noindent
$\phantom{{}^\dag{}\;}\ $E-mail: \it$^1$kyriakosnj20@gmail.com,
$^2$vaneeva@imath.kiev.ua,
$^3$christod@ucy.ac.cy
\par}

{\vspace{5mm}\par\noindent\hspace*{8mm}\parbox{146mm}{\small
 Lie symmetries of $K(m,n)$ equations with time-dependent coefficients are classified. Group classification is presented up to widest possible equivalence groups, the usual equivalence group of the whole class for the general case and conditional equivalence groups for special values of the exponents $m$ and $n$.
 Examples on reduction of  $K(m,n)$ equations (with  initial and boundary conditions)   to nonlinear ordinary differential equations (with initial conditions) are presented.
}\par\vspace{3mm}}


\section{Introduction}

In order to understand the role of nonlinear dispersion in the formation of patterns in liquid drops, Rosenau and Hyman \cite{rosenau93} introduced a generalization of the KdV equation of the form
\[
u_t+\varepsilon(u^m)_x+\left(u^n\right)_{xxx}=0,
\]
where $\varepsilon =\pm 1$.
Such equations, that are known as $K(m,n)$ equations, have the property for certain values of $m$ and $n$ their solitary wave solutions are of compact support. In other words, they vanish identically outside a finite core region. Further study followed in the references \cite{rosenau94,rosenau97,rosenau00,rosenau05}.

Here we consider a class of variable coefficient $K(m,n)$ equations of the form
\begin{equation}\label{3rdf(t)}
u_t+\varepsilon(u^m)_x+f(t)\left(u^n\right)_{xxx}=0,
\end{equation}
where $f$ is an arbitrary nonvanishing function of the variable $t$, $n$ and $m$ are arbitrary constants with $n\neq0$, and $\varepsilon=\pm1$.
 Note that the more general class (appeared, e.g., in \cite{Yin&lai&qing2009}) of the form
\begin{equation}\label{vcKmn}
u_t+g(t)(u^m)_x+f(t)(u^n)_{xxx}=0,\quad fn\neq0,
\end{equation}
reduces to class~\eqref{3rdf(t)} via the transformation
$
\tilde t=\varepsilon\int\!g(t){\rm d}t, $ $\tilde x=x,$ $\tilde u=u.
$ This transformation maps the
class~(\ref{vcKmn}) into its subclass (\ref{3rdf(t)}), where $\tilde
f=\varepsilon f/g$.
This is why without loss of generality it is sufficient to study  class~\eqref{3rdf(t)},
since all results on exact solutions, symmetries, conservation laws, etc.
for class~\eqref{vcKmn} can be derived form those
obtained for~\eqref{3rdf(t)} using the above transformation.

Lie symmetries have already been classified for many classes of constant coefficient
partial differential equations (PDEs) and for many
classes of PDEs involving functions with a range of
forms. Typically, extra symmetries exist for particular
forms of these functions. The classical method of finding Lie symmetries
is first to find infinitesimal transformations, with the benefit of linearization, and then
to extend these to groups of finite transformations. This method is easy to apply and
well-established in the last decades \cite{Bluman&Anco2002,Bluman89,Ibragimov99,Olver1986,Ovsiannikov1982}. This leads to a continuing interest in finding
exact solutions to nonlinear equations using Lie symmetries.

In the present paper, we carry out the Lie group classification for the class (\ref{3rdf(t)}).
All point transformations that link equations from the class are described. Firstly we find equivalence group of the entire class and then derive three its subclasses that have nontrivial conditional equivalence groups.
 The obtained Lie symmetries are employed also to a specific boundary value problem.

\section{Equivalence Transformations}

\looseness=-1
If two PDEs are connected by a point  transformation then these equations are called {\it similar}~\cite{Ovsiannikov1982} (it is possible also to consider a similarity up to  contact transformations). Similar PDEs have similar sets of solutions, symmetries, conservation laws and other related information.
Therefore, the important problem is the study of point transformations linking equations from a given class of PDEs. Such transformations are called
admissible~\cite{Popovych&Kunzinger&Eshraghi2010} (or form-preserving~\cite{Kingston&Sophocleous1998}) ones.
Admissible transformations that
preserve the differential structure of the class and transform only its arbitrary elements are called equivalence transformations and form a group.
Notions of different kinds of equivalence group can be found, e.g., in~\cite{vane2012b}.

The results on admissible transformations for equations from  the class (\ref{3rdf(t)}) are given in the following theorems. We exclude linear equations, i.e., equations with $(n,m)\in\{(1,0),(1,1)\}$, from the consideration. The proofs of these theorems are omitted. The detailed procedure of how to construct equivalence transformation (or point transformations in general) can be found, for example, in \cite{VPS2009,vane2012b}.

\begin{theorem}
The usual equivalence group~$G^{\sim}$ of the
class~\eqref{3rdf(t)} is formed by the transformations
\[
\begin{array}{c}
\tilde t=\pm\delta_1\delta_3^{\,1-m}t+\delta_0,\quad \tilde x=\delta_1x+\delta_2,  \quad
\tilde u=\delta_3u, \\[1ex]
\tilde f=\pm\delta_1^{2}\delta_3^{\,m-n} f, \quad \tilde\varepsilon=\pm\varepsilon,\quad
\tilde n=n, \quad \tilde m=m,
\end{array}
\]
where $\delta_j$, $j=0,1,2,3$, are arbitrary constants with $\delta_1\delta_3\not=0$.
\end{theorem}
It appears that if $(n,m)\in\{(n,0),(n,1),(1,2)\}$, then there exist nontrivial conditional equivalence groups of the class~\eqref{3rdf(t)} that are wider than $G^\sim,$ namely the following assertions are true.
\begin{theorem}
The class~\eqref{3rdf(t)} with $m=0$,
\begin{gather}\label{eq_m0}
u_t+f(t)\left(u^n\right)_{xxx}=0,
\end{gather}
admits usual equivalence group $G^{\sim}_{(n,0)}$ consisting of the transformations
\[
\begin{array}{c}
\tilde t=T(t),\quad \tilde x=\delta_1x+\delta_2,  \quad
\tilde u= \delta_3u, \quad
\tilde f=\dfrac{\delta_1^3\delta_3^{1-n}}{T_t}\, f, \quad
\tilde n=n,
\end{array}
\]
where $\delta_j$, $j=1,2,3$, are arbitrary constants with $\delta_1\delta_3\not=0$, $T(t)$ is an arbitrary smooth function with $T_t\neq0.$
\end{theorem}
\begin{theorem}
The generalized equivalence group $G^{\sim}_{(n,1)}$ of the class~\eqref{3rdf(t)} with $m=1$,
\begin{gather}\label{eq_m1}
u_t+\varepsilon u_x+f(t)\left(u^n\right)_{xxx}=0,
\end{gather}
comprises the transformations
\[
\begin{array}{c}
\tilde t=T(t),\quad \tilde x=\delta_1 (x-\varepsilon t)\pm\varepsilon T(t)+\delta_2,  \quad
\tilde u= \delta_3u, \\[1ex]
\tilde f=\dfrac{\delta_1^3\delta_3^{1-n}}{T_t}\, f, \quad \tilde\varepsilon=\pm\varepsilon,\quad
\tilde n=n,
\end{array}
\]
where $\delta_j$, $j=1,2,3$, are arbitrary constants with $\delta_1\delta_3\not=0$, $T(t)$ is an arbitrary smooth function with $T_t\neq0.$
\end{theorem}
\begin{theorem}
The generalized equivalence group~$
G^{\sim}_{(1,2)}$ of the class,
\begin{equation}\label{Eq_GenBurgers_n2}
u_t+\varepsilon(u^2)_x+f(t)u_{xxx}=0,
\end{equation}
 consists of  the transformations
\[
\begin{array}{c}
\tilde t=\dfrac{\alpha t+\beta}{\gamma t+\delta},\quad \tilde x=\dfrac{\kappa x +\mu_1t+\mu_0}{\gamma t+\delta},\quad
\tilde u=\pm\dfrac{2\varepsilon\kappa(\gamma t+\delta)u-\kappa\gamma x+\mu_1\delta-\mu_0\gamma}{2\varepsilon(\alpha\delta-\beta\gamma)}, \\[2ex]
\tilde \varepsilon=\pm\varepsilon \quad \mbox{\rm and} \quad
\tilde f=\dfrac{\kappa^3}{\alpha\delta-\beta\gamma}\dfrac{f}{\gamma t+\delta},
\end{array}
\]
where
 $\alpha, \beta, \gamma, \delta,\mu_1, \mu_0,$ and $ \kappa$ are constants defined up to a nonzero multiplier,
 $\kappa(\alpha\delta-\beta\gamma)\neq0$.
\end{theorem}
The equations  from the class~\eqref{eq_m1} can be reduced to ones (with tilded variables) from the class~\eqref{eq_m0} by the additional equivalence transformation
\[
\tilde t=t,\quad \tilde x=x-\varepsilon t,\quad \tilde u=u.
\]
Therefore, the case $m=1$ being equivalent to the case $m=0$ will be excluded from the classification list.
\section{Lie Symmetries}

We perform the group classification of class~\eqref{3rdf(t)} within the framework of
the classical Lie approach~\cite{Olver1986,Ovsiannikov1982,Bluman89,Ibragimov99}.
We search for operators of the form
\[
\Gamma=\tau(t,x,u)\partial_t+\xi(t,x,u)\partial_x+\eta(t,x,u)\partial_u
\]
which generate one-parameter groups of point-symmetry transformations of an equation from class~\eqref{3rdf(t)}.
Any such vector field,~$\Gamma$, satisfies the infinitesimal invariance criterion, i.e.,
the action of the third prolongation,~$\Gamma^{(3)}$, of the operator~$\Gamma$ on equation~\eqref{3rdf(t)}
results in the condition being an identity for all solutions of this equation. That is, we require that
\begin{gather}\label{conditionf(t)3}\arraycolsep=0ex
\begin{array}{l}
\Gamma^{(3)}\Big[u_t+\varepsilon mu^{m-1}u_x+
 nf(t)\big(u^{n-1}u_{xxx}+3(n-1)u^{n-2}u_xu_{xx}\\[1ex]\qquad\qquad+(n-1)(n-2)u^{n-3}u_x^3\big)
\Big]=0
\end{array}
\end{gather}
identically, modulo equation~\eqref{3rdf(t)}.

Equation \eqref{conditionf(t)3} is an identity in the variables$~u_x,~u_{xx},~u_{tx},~u_{xxx}~$and$~u_{txx}$.
Coefficients of different powers of these variables, which must be equal to
zero, lead to the determining equations on the coefficients $\tau$, $\xi$ and $\eta$.
Firstly, we use the general results on point transformations between evolution equations~\cite{Kingston&Sophocleous1998} which simplify the forms of the coefficient functions. Specifically, we have
$\tau =\tau(t)~$and$~\xi =\xi(t,x)$. Since the procedure is quite straightforward
\cite{Olver1986,Ovsiannikov1982,Bluman89,Ibragimov99}, we omit the detailed analysis. However we point out that the classification of Lie symmetries is complete.

The coefficient of $u_{xxx}$ gives the equation
\[
\left[f_t\tau+f(\tau_t-3\xi_x)\right]u+(n-1)f\eta=0
\]
from which we deduce that the analysis needs to be split in two exclusive cases: $n\neq1$ and $n=1$.

{\bf I.} If $n\neq 1$, the coefficient function $\eta$ takes the form
\[
\eta=-\frac{[f_t\tau+f(\tau_t-3\xi_x)]u}{(n-1)f}.
\]
The coefficients of  $u_{xx}$ (or $u_{x}^2$), $u_{x}$ and the term independent of derivatives in \eqref{conditionf(t)3} produce the following determining equations, respectively,
\begin{gather*}\label{ux2f(t)}
n(2n+1)f\xi_{xx}=0,\\\arraycolsep=0ex
\begin{array}{l}\label{uxf(t)}
\varepsilon m\left[(m-n)f\tau_t-(3m-n-2)f\xi_x+(m-1)f_t\tau\right]u^m\\[1ex]
\quad-n(8n+1)f^2\xi_{xxx}u^n+(n-1)f\xi_tu=0,
\end{array}\\
\label{ux^0f(t)}
3\varepsilon mf^2\xi_{xx}u^{m}+3nf^3\xi_{xxxx}u^{n}-\left[f^2\tau_{tt}+ff_t\tau_t+ff_{tt}\tau-f_t^2\tau-3 f^2\xi_{tx}\right]u=0.
\end{gather*}
Solution of the above determining equations leads to the forms of $\tau(t)$, $\xi(t,x)$ and  $f(t)$. The forms of $f(t)$ are determined using {\it the method of furcate split} suggested in~\cite{Popovych&Ivanova2004}.
Lie symmetries according to the forms of $f(t)$ are tabulated in Table~1.

\begin{table}[h!]\small
\renewcommand{\arraystretch}{1.7}
\begin{center}
\label{table8f(t)}
\caption{\small Classification of the equations \eqref{3rdf(t)} with $n\neq1$. }
\begin{tabular}{|c|c|c|c|l|}
\hline
no.&$n$&$m$&$f(t)$&\hfil Basis of $A^{\max}$ \\
\hline
1&$\forall$&$\forall$&$\forall$&$\partial_x$\\
\hline
2&$\forall$&$\frac{n+2}3$&$\forall$&$\partial_x,\,x\partial_x+\frac{3}{n-1}u\partial_u$\\
\hline
3&$\forall$&$0$&$1$&$ \partial_t,\,\partial_x,\,x\partial_x+\frac{3}{n-1}u\partial_u,\,3t\partial_t+x\partial_x$\\
\hline
4&$-\frac12$&$0$&$1$&$ \partial_t,\,\partial_x,\,x\partial_x-2u\partial_u,\,3t\partial_t+x\partial_x,\,x^2\partial_x-4xu\partial_u$\\
\hline
5&$\forall$&$\forall$&$1$&$\partial_t,\,\partial_x,\,(3m\!-\!n\!-\!2)t\partial_t+(m\!-\!n)x\partial_x-2u\partial_u$\\
\hline
$6_a$&$-\frac12$&$-\frac12$&$1$&$\partial_t,\,\partial_x,\,3t\partial_t+2u\partial_u,$\\
&&&&$\sin x\,\partial_x-2\cos x\, u\partial_u,\,\cos x\,\partial_x+2\sin x\, u\partial_u$\\
\hline
$6_b$&$-\frac12$&$-\frac12$&$1$&$\partial_t,\,\partial_x,\,3t\partial_t+2u\partial_u,\,e^x\partial_x-2e^xu\partial_u,\,e^{-x}\partial_x+2e^{-x}u\partial_u$\\
\hline
7&$\forall$&$\forall$&$t^k$&$\partial_x,\,(3m\!-\!n\!-\!2)t\partial_t+(km\!-k\!+m\!-\!n)x\partial_x+(k\!-\!2)u\partial_u$\\
\hline
8&$\forall$&$\frac{n+2}3$&$t^2$&$\partial_x,\,x\partial_x+\frac{3}{n-1}u\partial_u,\,t\partial_t+x\partial_x$\\
\hline
9&$\forall$&$\forall$&$e^{t}$&$\partial_x,\,(3m\!-\!n\!-\!2)\partial_t+(m\!-\!1)x\partial_x+u\partial_u$\\
\hline
\end{tabular}
\end{center}
Here $k$ is an arbitrary nonzero constant. In Cases $6_a$ and $6_b$ $\varepsilon=1$ and $\varepsilon=-1$, respectively.
\end{table}
\begin{remark}
All cases presented in Table~1 except Cases 3 and 4 are classified up to $G^{\sim}$-equivalence.
For Cases 3 and 4, where $m=0$, we
used the equivalence group $G^{\sim}_{(n,0)}$ that is wider than $G^{\sim}$. Thus, the equation~\eqref{eq_m0} with $n\neq-1/2$ admits the four-dimensional Lie symmetry algebra with the basis operators
\[\frac1{f(t)}\partial_t, \quad\partial_x,\quad x\partial_x+\frac{3}{n-1}u\partial_u,\quad 3\frac{\int\!{f(t){\rm d}t}}{f(t)}\partial_t+x\partial_x
\] irrespectively of the form of the function~$f$.  Here and throughout the paper an integral with respect to~$t$ should be interpreted as a fixed antiderivative. If $n=-1/2$, then the Lie symmetry algebra of the equation~\eqref{eq_m0} is five-dimensional
 spanned by the above operators and the additional operator $x^2\partial_x-4xu\partial_u$.
Using the equivalence transformation $\tilde t=\int\!f(t){\rm d}t$, $\tilde x=x$, $\tilde u=u$ from the group $G^{\sim}_{(n,0)}$ we reduce these cases to ones with $f=1$ (cf., Cases~3 and 4 of Table~1).
\end{remark}

{\bf II.} If $n=1$, the coefficient of $u_xu_{xx}$ implies that $\eta_{uu}=0$, and so
$
\eta=\eta^1(t,x)u+\eta^0(t,x).
$
We use the fact that $\tau=\tau(t)$, $\xi=\xi(t,x)$ and the above form for $\eta$ and from \eqref{conditionf(t)3} we obtain the following determining equations
\begin{gather*}
f_t\tau+f(\tau_t-3\xi_x)=0,\quad
\eta^1_x-\xi_{xx}=0,\\
\varepsilon m\!\left(\!\tau_t-\xi_x+(m-1)\eta^1\!\right)\!u^{m}\!+\!\varepsilon m(m-1)\eta^0u^{m-1}\!\!+\!(3f\eta^1_{xx}-\xi_t-f\xi_{xxx})u=0,
\\
\varepsilon m\eta^1_xu^{m+1}+\varepsilon m\eta^0_xu^m+(\eta^1_t+f\eta^1_{xxx})u^2+(\eta^0_t+f\eta^0_{xxx})u=0.
\end{gather*}
We solve the above system  and adduce the results in Table~2. It is worthy to note that
the group classification problem for the class of equations
$u_t+u^{\bar m}u_x+\bar f(t)u_{xxx}=0$ with $\bar m\bar f\ne0$, that are  similar to equations of the form~\eqref{3rdf(t)} with $n=1$,
was carried out in~\cite{Popovych&Vaneeva2010,VPCS2013}.
\begin{table}[h!]\small
\renewcommand{\arraystretch}{1.7}
\begin{center}
\label{tablef(t)}
\caption{\small Classification of the class \eqref{3rdf(t)} with $n=1$.  }
\begin{tabular}{|c|c|l|}
\hline
no.&$f(t)$&\hfil Basis of $A^{\max}$ \\
\hline
\multicolumn{3}{|c|}{$m\neq2$}\\
\hline
1&$\forall$&$\partial_x$\\
\hline
2&1&$\partial_t,\,\partial_x,\,3(m-1)t\partial_t+(m-1)x\partial_x-2u\partial_u$\\
\hline
3&$t^k$&$\partial_x,\,3(m-1)t\partial_t+(m-1)(k+1)x\partial_x+(k-2)u\partial_u$\\
\hline
4&$e^{t}$&$\partial_x,\,3(m-1)\partial_t+(m-1)x\partial_x+u\partial_u$\\
\hline
\multicolumn{3}{|c|}{$m=2$}\\
\hline
5&$\forall$&$\partial_x,\,2\varepsilon t\partial_x+\partial_u$\\
\hline
6&1&$\partial_t,\,2\varepsilon t\partial_x+\partial_u,\,\partial_x,\,3t\partial_t+x\partial_x-2u\partial_u$\\
\hline
7&$t^k$&$\partial_x,\,2\varepsilon t\partial_x+\partial_u,\,3t\partial_t+(k+1)x\partial_x+(k-2)u\partial_u$\\
\hline
8&$e^{t}$&$\partial_x,\,2\varepsilon t\partial_x+\partial_u,\,3\partial_t+x\partial_x+u\partial_u$\\
\hline
9&$e^{k \arctan t}\sqrt{t^2+1}$&$\partial_x,\,2\varepsilon t\partial_x+\partial_u,\,6\varepsilon (t^2+1)\partial_t+2\varepsilon(3t+k )x\partial_x+(2\varepsilon(k -3t)u+3x)\partial_u$\\
\hline
\end{tabular}
\end{center}
Here $k$ is an arbitrary constant satisfying the following constraints: $k\neq0$ in Case 3, $k\neq0,1$ and $k\geqslant1/2\bmod G^{\sim}_{(1,2)}$ in Case 7, $k\geqslant0\bmod G^{\sim}_{(1,2)}$ in Case 9.
\end{table}
\begin{remark}
The group classification of the class~\eqref{3rdf(t)} with $n=1$ and $m\neq2$ is performed up to $G^\sim$-equivalence. For the classification of Lie symmetries
of the equations~\eqref{3rdf(t)} with $n=1$ and $m=2$ we used the wider conditional equivalence group $G^{\sim}_{(1,2)}$. Since transformations from the  group $G^{\sim}_{(1,2)}$ are quite complicated, we adduce also the additional cases of Lie symmetry extensions of equations~\eqref{3rdf(t)} with $n=1$ and $m=2$ that are inequivalent with respect to the group $G^{\sim}$ to Cases 6--9 of Table~2.
\medskip

1. $\displaystyle f=(t+\beta)^kt^{1-k}$,
$k\ne0,1,\,\beta\neq0$:\quad$\langle\partial_x,\,2\varepsilon t\partial_x+\partial_u,\ \Gamma_3\rangle$,\quad where
\[\Gamma_3= 6\varepsilon t(t+\beta)\partial_{t}+
 2\varepsilon \left(3t+\beta(2-k)\right)  x\partial_{x}+\left[3x-2\varepsilon(3t+\beta(k+1)) u\right]\!\partial_{ u};
\]

2. $\displaystyle f=te^{\frac1t}$:\quad
$\langle\partial_x,\,2\varepsilon t\partial_x+\partial_u,\ 6\varepsilon t^2\partial_{t}+2\varepsilon(3t-1)x\partial_x+(3x-2\varepsilon(3t+2) u)\partial_{ u}\rangle$;

\medskip

3. $f=t$: \quad
$\langle\partial_x,\,2\varepsilon t\partial_x+\partial_u,\ 3t\partial_{ t}+2x\partial_{ x}-u\partial_{ u},\ 2\varepsilon t^2\partial_{ t}+  2\varepsilon tx\partial_{ x}+(x- 2\varepsilon tu)\partial_{ u}\rangle$.

\medskip

From the first sight it looks like the counterpart to Case 9 of Table~1 is missed. At the same time it appears that the function  $ f=\lambda\exp\!\Big({k\arctan\!\frac{\alpha t+\beta}{\gamma t+\delta}}\Big)$ locally coincides with the function $\check f=\check\lambda\exp( k\arctan(\check\alpha t+\check\beta))$, see~\cite{PPV} for details.

\end{remark}

The primary use of Lie symmetries is to obtain a reduction of variables. Similarity variables appear as first integrals
 of the characteristic system
\[
\frac{{\rm d}t}{\tau}=\frac{{\rm d}x}{\xi}=\frac{{\rm d}u}{\eta}.
\]
Here we can reduce a PDE in two independent variables into an ordinary differential equation (ODE) using a one-dimensional subalgebra of Lie symmetry algebra.
Reductions could be obtained from any symmetry which is an arbitrary
linear combination
$
\sum_{i=1}^s a_i\Gamma_i,
$
where $s$ is the number of basis operators of maximal Lie symmetry algebra of the given PDE.
To ensure that a minimal complete set of reductions is obtained from the Lie
symmetries of  equation~(\ref{3rdf(t)}), we construct the so-called optimal system of one-dimensional subalgebras. Ovsiannikov \cite{Ovsiannikov1982} proved  that the optimal system of solutions consists of solutions that are invariant with respect to all proper inequivalent subalgebras of the symmetry algebra. More detail about construction of optimal sets of subalgebras can be found in~\cite{Ovsiannikov1982,Olver1986}.

As an example for a reduction into an ordinary differential equation, we consider Case 7 of Table 2 which corresponds to the variable coefficient KdV equation
\begin{equation}\label{case7}
u_t+\varepsilon(u^2)_x+t^ku_{xxx}=0
\end{equation}
that admits the three-dimensional Lie symmetry algebra
\[
\Gamma_1=\partial_x,~~\Gamma_2=2\varepsilon t\partial_x+\partial_u,~~\Gamma_3 =3t\partial_t+(k+1)x\partial_x+(k-2)u\partial_u.
\]
Depending on the value of $k$
an optimal system of one-dimensional subalgebras of this Lie symmetry algebra consists of the  subalgebras
\begin{gather*}\arraycolsep=0ex
\begin{array}{ll}
\langle \Gamma_1 \rangle,~~\langle \Gamma_2+\sigma\Gamma_1 \rangle,~~\langle \Gamma_3  \rangle&
\quad\mbox{if}\quad k\neq-1,2\\[1ex]
\langle \Gamma_1 \rangle,~~\langle \Gamma_2+\sigma\Gamma_1 \rangle,~~\langle \Gamma_3+a\Gamma_1  \rangle
&\quad\mbox{if}\quad k=-1\\[1ex]
\langle \Gamma_1 \rangle,~~\langle \Gamma_2+\sigma\Gamma_1 \rangle,~~\langle \Gamma_3+a\Gamma_2  \rangle &\quad\mbox{if}\quad k=2.
\end{array}
\end{gather*}
Here $\sigma\in\{-1,0,1\},$ $a\in\mathbb{R}.$

Reductions associated with the subalgebra $\langle\Gamma_1\rangle$ are not considered
since they lead to constant solutions only. The ansatz constructed with the subalgebra $\langle\Gamma_2+\sigma\Gamma_1\rangle$ has the form
$u=\dfrac{x}{2\varepsilon t+\sigma}+\phi(\omega)$ with the similarity variable $\omega=t$. This ansatz reduces equation~\eqref{case7} to the ODE
$(2\varepsilon\omega+\sigma)\phi_{\omega}+2\varepsilon\phi=0$ whose general solution is $\phi=\dfrac{c_1}{2\varepsilon \omega+\sigma}$, where $c_1$ is an arbitrary constant. The corresponding solution of~\eqref{case7} takes the form
\[
u=\frac{x+c_1}{2\varepsilon t+\sigma}.
\] It is fair to note that this solution satisfies equations of the form~\eqref{Eq_GenBurgers_n2} for arbitrary $f$.
Other reductions depend on the value of the exponent $k$. We adduce the ansatzes together with the corresponding reduced equations.
\begin{gather*}
k\neq -1,2.\quad\langle\Gamma_3\rangle\colon u=t^{\frac{k-2}{3}}\phi(\omega),~~~\omega=xt^{-\frac{k+1}{3}}\\
3\phi_{\omega\omega\omega}+6\varepsilon \phi \phi_{\omega}-(k+1)\phi_{\omega}\omega+(k-2)\phi=0.
\end{gather*}
\begin{gather*}
k= -1.\quad\langle\Gamma_3+a\Gamma_1\rangle\colon~u=\frac1t\phi(\omega),~~~\omega=x-\frac{ a}{3}\ln t\\
3\phi_{\omega\omega\omega}+6\varepsilon \phi \phi_{\omega}-a\phi_{\omega}-3\phi=0.
\end{gather*}
\begin{gather*}
k=2.\quad \langle\Gamma_3+a\Gamma_2\rangle\colon~u=\frac{a}{3}\ln t+\phi(\omega),~~~\omega=\frac{x}{t}-\frac{2a\varepsilon}{3}\ln t\\
3\phi_{\omega\omega\omega}+6\varepsilon \phi \phi_{\omega}-3\phi_{\omega}\omega-2a\varepsilon\phi_{\omega}+a=0.
\end{gather*}
We note that the latter two cases are equivalent. Indeed,  the equation
\[u_t+\varepsilon(u^2)_x+t^2u_{xxx}=0\] is mapped to the equation \[\tilde u_{\tilde t}+\varepsilon{(\tilde u)^2}_{\tilde x}+{\tilde t}^{-1}{\tilde u}_{\tilde x\tilde x\tilde x}=0\]
by the following transformation from the group~$G^\sim_{(1,2)}$\[\tilde t=\frac1t, \quad\tilde x=-\frac xt,\quad\tilde u=\frac{2\varepsilon tu-x}{2\varepsilon}.\]

\section{Application of Lie symmetries to a Boundary Value Problem}
There exist several techniques that use Lie symmetries in reduction of boundary-value problems (BVPs) for PDEs to those for ODEs.
The classical method suggested in~\cite{Bluman&Cole1969,Bluman1974}
is to require that both equation and boundary conditions are left invariant under the one-parameter Lie group
of infinitesimal transformations. Of course the infinitesimal approach is usually applied, i.e., a basis of operators of Lie invariance algebras is used instead of finite transformations from the corresponding Lie symmetry group  (see, e.g.,~\cite[Section 4.4]{Bluman&Anco2002}).
Firstly, the symmetries of a PDE should be derived and then the boundary conditions should be checked to determine whether they are also invariant under the action of the generators of the symmetries found. In the case of a positive answer the BVP for the PDE was reduced to a BVP for an ODE. Using this technique a number of boundary-value problems were solved successfully  (see, e.g.,~\cite{Sophocleous&OHara&Leach2011a,Sophocleous&OHara&Leach2011b,VSL}).

Consider the initial and boundary value problem
\begin{gather}\label{BV_Kmn}\arraycolsep=0ex
u_t+\varepsilon(u^m)_x+t^{ k }(u^n)_{xxx}=0,\quad t>0,\quad x>0, \\[1ex]\label{BV_Kmn_bc}
\begin{array}{l}
u(x,0)=0, \quad x>0,\\[0.5ex]
u(0,t)=q(t),\quad
u_x(0,t)=0, \quad
u_{xx}(0,t)=0,\quad t>0.
\end{array}
\end{gather}
We look for a nonconstant solution using the ``direct'' approach suggested by Blu\-man~\cite{Bluman&Cole1969,Bluman&Anco2002}.

The Lie symmetries for the variable coefficient equation  (\ref{BV_Kmn}) are derived in the previous section and now we check which of these symmetries leave the associated initial and boundary conditions invariant. For equation~\eqref{BV_Kmn} we should consider a general symmetry operator of the form
\begin{equation}\label{general_symmetry}
\Gamma =\sum_{i=1}^s\alpha_i\Gamma_i,
\end{equation}
where $s$ is the number of basis operators of its maximal Lie symmetry algebra  and  $\alpha_i,~i = 1,\dots,s$, are constants to be determined.

Equation~\eqref{BV_Kmn} admits for arbitrary $n$, $m$ and $ k $ a two-dimensional Lie symmetry algebra with the basis operators
\[\Gamma_1=\partial_x,\quad\Gamma_2=(3m-n-2)t\partial_t+( k  m- k +m-n)x\partial_x+( k -2)u\partial_u.\]
 The general symmetry (\ref{general_symmetry}) takes the form
\[
\Gamma=\alpha_1\partial_x+\alpha_2\left [(3m-n-2)t\partial_t+( k  m- k +m-n)x\partial_x+( k -2)u\partial_u \right ].
\]
Application of $\Gamma$ to the first boundary condition
$x=0,~u(t,0)=q(t)$ gives
$
\alpha_1=0 $ and $q(t)=\gamma t^{\frac{ k -2}{3m-n-2}},
$ $m\neq\frac{n+2}3.$
Using the second extension of $\Gamma$,
\begin{gather*}
\Gamma^{(2)}=3nt\partial_t+( k +1)n x\partial_x+( k -2) u\partial_u+( k -n k -n-2)u_x\partial_{u_x}\\+( k -2n k -2n-2)u_{xx}\partial_{u_{xx}},
\end{gather*}
where the unused terms have been ignored, it can be shown that it leaves invariant the initial condition and the remaining two boundary conditions of~\eqref{BV_Kmn_bc}. Finally, symmetry $\Gamma$ produces the transformation
\begin{equation}\label{trr}
u=t^{\frac{ k -2}{3m-n-2}}\phi(\omega),~~~\omega=xt^{-\frac{ k  m- k +m-n}{3m-n-2}},
\end{equation}
which reduces the problem~\eqref{BV_Kmn}--\eqref{BV_Kmn_bc} into
\begin{gather}\arraycolsep=0ex
\begin{array}{l}
 (\phi^n)'''+\varepsilon(\phi^m)'-\dfrac{ k  m- k +m-n}{3m-n-2}\omega\phi'+\dfrac{ k -2}{3m-n-2}\phi=0, \\[2ex]
\phi(0)=\gamma, \qquad
 \phi'(0)=0, \qquad
 \phi''(0)=0.
 \end{array}
\end{gather}
The latter IVP can be solved numerically and then the solution of IBVP~\eqref{BV_Kmn}--\eqref{BV_Kmn_bc} can be recovered using the transformation~\eqref{trr}. See~\cite{VPCS2013}, where a similar problem for generalized KdV equations was solved successfully using finite difference method, for details.

\section*{Acknowledgements}
KC is grateful to the University of Cyprus for financial support. OV expresses the gratitude to the hospitality shown by the University of Cyprus during her visits. The authors also would like to thank Prof. Roman Popovych for useful comments.


 \end{document}